\documentclass[10pt,prl,aps,twocolumn,superscriptaddress,floatfix,showpacs,nobalancelastpage]{revtex4}

\pdfoutput=1

\usepackage{graphicx}
\usepackage{amssymb}
\usepackage{bm}
\usepackage{bbold}
\usepackage[toc,page]{appendix} 

\usepackage{color}

\begin{document}

\title{SU(2)$_1$ chiral edge modes of a critical spin liquid}

\author{Didier Poilblanc}
\affiliation{Laboratoire de Physique Th\'eorique, C.N.R.S. and Universit\'e de Toulouse, 31062 Toulouse, France}

\author{Norbert \surname{Schuch}}
\affiliation{Max-Planck-Institut f{\"{u}}r Quantenoptik, Hans-Kopfermann-Str.\ 1, D-85748 Garching, Germany}
\affiliation{Institut f\"ur Quanteninformation, RWTH Aachen University, D-52056 Aachen, Germany}

\author{Ian Affleck}
\affiliation{Department of Physics and Astronomy, University of British Columbia, Vancouver, B.C., Canada, V6T1Z1} 

\date{\today}

\begin{abstract}
Protected chiral edge modes are a well-known signature of topologically ordered phases like the Fractional Quantum Hall States.
Recently, using the framework of projected entangled pair states (PEPS) on the square lattice, we constructed a family of 
chiral  Resonating Valence Bond states with $\mathbb{Z}_2$ gauge symmetry.  Here we revisit and analyze in full details the properties of the edge modes as given by their
 Entanglement Spectra on a cylinder. Surprisingly, we show that the latter can be well described  by
 a chiral SU(2)$_1$ Conformal Field Theory, as for the $\nu=1/2$ (bosonic) gapped Laughlin state, although our numerical data suggest a critical bulk 
 compatible with an emergent $U(1)$ gauge symmetry.
 We propose that our family of PEPS may physically describe a boundary between a chiral topological phase and a trivial phase. 
  \end{abstract}
\maketitle

{\it Introduction} -- 
Fractional Quantum Hall States (FQHS) are remarkable incompressible phases of electronic matter exhibiting 
chiral edge modes~\cite{wen} which are protected by the long-range topological order of the bulk. 
Famous example of FQHS are the abelian bosonic Laughlin state~\cite{laughlin} and the non-Abelian Moore-Read~\cite{moore} 
and Read-Rezayi~\cite{read} states.
Remarkably, their edge physics can be described by chiral $SU(2)_k$ 
Wess-Zumino-Witten  (WZW)~\cite{CFT} Conformal Field Theory (CFT), where $k=1,2$ and $3$ respectively.
In the simplest Abelian case of interest here, the $SU(2)_1$ edge theory simply corresponds to chiral free bosons.

Topological order also exists in quantum spin systems,  
e.g. in Kitaev's Toric Code~\cite{TC} or in short-range (non-chiral) $\mathbb Z_2$ spin liquids
as the spin-1/2 Resonating Valence Bond (RVB) state -- an equal weight superposition of nearest-neighbor (NN) singlet 
coverings~\cite{RVB0} -- on non-bipartite lattices~\cite{wildeboer,yangfan}. In contrast, 
on bipartite lattices like the square lattice, generic dimer or RVB liquids have 
{\it power law decay}  of dimer correlations~\cite{albuquerque,tang}. 
In all cases, the Projected Entangled Paired States (PEPS) framework~\cite{frank1,cirac} is a practical formalism to
gain a comprehensive understanding of RVB wave functions~\cite{frank2,norbert,didier,wang}.  

Long ago, topological chiral spin liquids (CSL), spin analogs of the Laughlin state, have been proposed in 
seminal work by Kalmeyer and Laughlin~\cite{KL} and Wen, Wilczek and Zee~\cite{WWZ}, and more recently, parent Hamiltonians
have been constructed for such states~\cite{TG,nielsen}. 
In addition, some quantum antiferromagnets with broken time-reversal symmetry and {\it local interactions} have been shown to host
topological CSL of the same class as the bosonic Laughlin state~\cite{bauer,nielsen}. 

Efforts to construct chiral topological states on the lattice within the PEPS framework have been undertaken recently.
By Gutzwiller projecting two copies of {\it non-interacting}
Gaussian fermionic PEPS~\cite{shuo}, a chiral (critical) PEPS with topological order was constructed, from which
the edge CFT could be identified using the PEPS bulk-edge correspondence~\cite{cirac}.
It was also shown that a simple two-dimensional family of SU(2)-invariant PEPS describe topological chiral spin liquids, although with algebraic (singlet) correlations~\cite{chiral_SL}. In this article, we focus in more details on  the properties of the edge states,
as defined from the Entanglement Spectrum~\cite{LiHaldane}, of such a critical CSL and provide evidence that they can be fully described  by a chiral SU(2)$_1$ Conformal Field Theory.

\begin{figure}
\begin{center}
\includegraphics[width=8cm,angle=0]{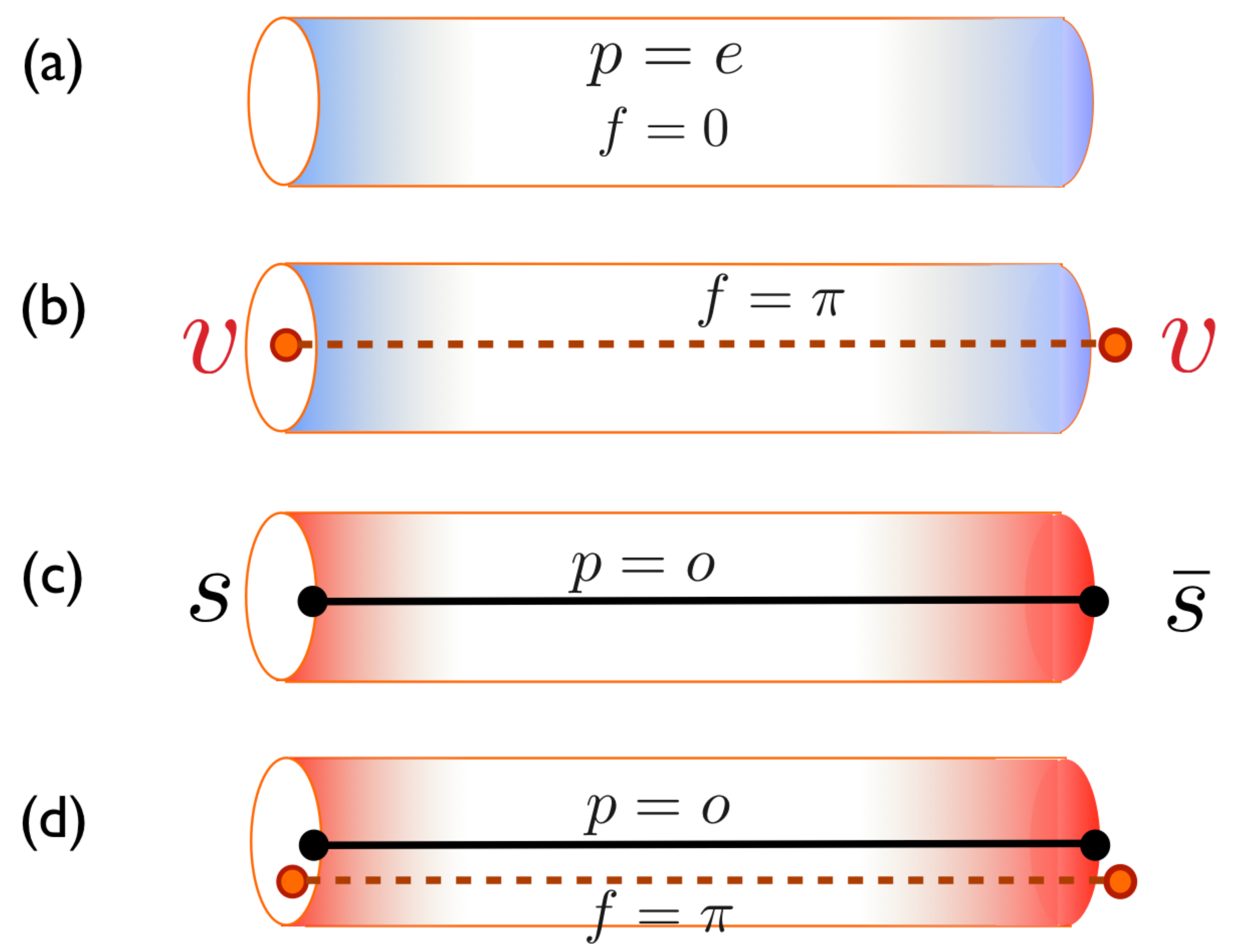}
\caption{The RVB state on infinite cylinders. The parity $p$ (even or odd) and the $\mathbb{Z}_2$-flux $f$ ($0$ or  $\pi$) topological invariant can be switched by 
inserting/removing ``spinon" (s) or ``vison" (v) lines through the cylinder (see Ref.~\cite{didier} for details) as in 
Kitaev's code~\cite{TC}. }
\label{Fig:sec}
\end{center}
\vskip -0.5cm
\end{figure}

{\it Chiral RVB} -- 
We start by reviewing briefly the construction of the CSL of Ref.~\onlinecite{chiral_SL} which is based 
on a deformation of the simple PEPS ansatz of the RVB state on the two-dimensional (2d) bipartite square lattice~\cite{didier,wang}.
The RVB PEPS ansatz is based on a tensor network
involving a unique rank-5 tensor $A$ assigned to each site of the square lattice. $A$ involves local physical spin-1/2 degrees of freedom (of dimension $d=2$) and virtual states on the four surrounding bonds belonging to the $1/2\oplus 0$ spin representation of dimension $D=3$~\cite{norbert,didier}. 
We modify the (non-zero) tensor elements of the real (critical) RVB state~\cite{didier,wang}  
as~:
\begin{equation}
A=\lambda_1 R_1 +\lambda_2 R_2 + i \lambda_{\rm chiral} I_c  ,
\label{Eq:tensor}
\end{equation}
where the $R_l$ and $I_c$ tensors transform, under point group operations, according to the 
$B_1$ ($d_{x^2-y^2}$-wave orbital symmetry) and $B_2$ ($d_{xy}$-wave orbital symmetry) 
irreducible representations (IRREPs) of the ($C_{4v}$) point group of the square lattice, respectively, 
and where $\lambda_1$, $\lambda_2$ and 
$\lambda_{\rm chiral}$ are free (real) parameters (see Appendix). Note that the gauge symmetry of the tensor $A$~\cite{ran}
is broken down from $U(1)$ to $\mathbb{Z}_2$ upon including the $R_2$ and/or $I_c$ tensors.
The resulting complex PEPS wavefunction is found to be 
a chiral RVB SL~\cite{chiral_SL}, 
\begin{equation}
|\Psi_{\rm Chiral}\rangle =|\Psi_{s}\rangle + i |\Psi_{g}\rangle .
\label{s+ig}
\end{equation}
whose real and imaginary components transform according to
the $A_1$ ($s$-wave orbital symmetry) and $A_2$ ($g$-wave orbital symmetry) IRREPs. 
Hence, under any reflection symmetry of $C_{4v}$,  
$|\Psi_{\rm chiral}\rangle$ transforms into $|\Psi^*_{\rm chiral}\rangle$, 
its time-reversed state (possibly up to a sign), which is indeed a necessary condition for a chiral spin liquid.
For $\lambda_2=\lambda_{\rm chiral}=0$, one recovers the (non-chiral) NN RVB state~\cite{norbert,didier}. 
$\lambda_2$ and $\lambda_{\rm chiral}$ introduce longer range (AB) singlet bonds via ``teleportation''
as in the non-chiral spin liquid of Ref.~\onlinecite{wang}.
Here we choose  $\lambda_1=\lambda_2=\lambda_{\rm chiral}(=1)$ for simplicity. 

{\it RDM \& Entanglement Hamiltonian} -- Using the standard procedure~\cite{cirac}, we have computed the Reduced Density Matrix (RDM) for an infinite cylinder cut into two (semi-infinite) halves.  
Following the procedure for the RVB state, one can construct {\it a priori} four sectors labelled by a pair of  parity/flux quantum number 
$\mu=(p,f)$. Even ($p=e$) and odd ($p=o$) parities as well as zero ($f=0$) or $\pi$ ($f=\pi$) $\mathbb{Z}_2$ (``vison") fluxes can be considered
as shown in Fig.~\ref{Fig:sec}. 
The right and left eigenvectors corresponding to the leading (largest) eigenvalues of the TM
in each sector $\mu$ 
can be viewed as edge operators $\sigma_R^{\mu}$ and $\sigma_L^{\mu}$ acting on 
the virtual $0\otimes 1/2$ indices of the edges.
One then obtains the RDM (up to an isometry) as $\rho^{(\mu)}=\sqrt{(\sigma_R^\mu)^\top}\sigma_L^\mu\sqrt{(\sigma_R^\mu)^\top}$
and the Entanglement Hamiltonian  $H_E^{(\mu)}$ implicitly defined by
$\rho^{(\mu)}=\exp{(-H_E^{(\mu)})}$. Hence, in this language, the RDM can be viewed as a thermal density matrix at ``temperature''
$1/\beta=1$.
The Entanglement Hamiltonian is block-diagonal and the blocks are labelled by the edge momentum $K$ and the (modulus of the) virtual edge spin component $S_z$. 

\begin{figure}
\begin{center}
\includegraphics[width=\columnwidth,angle=0]{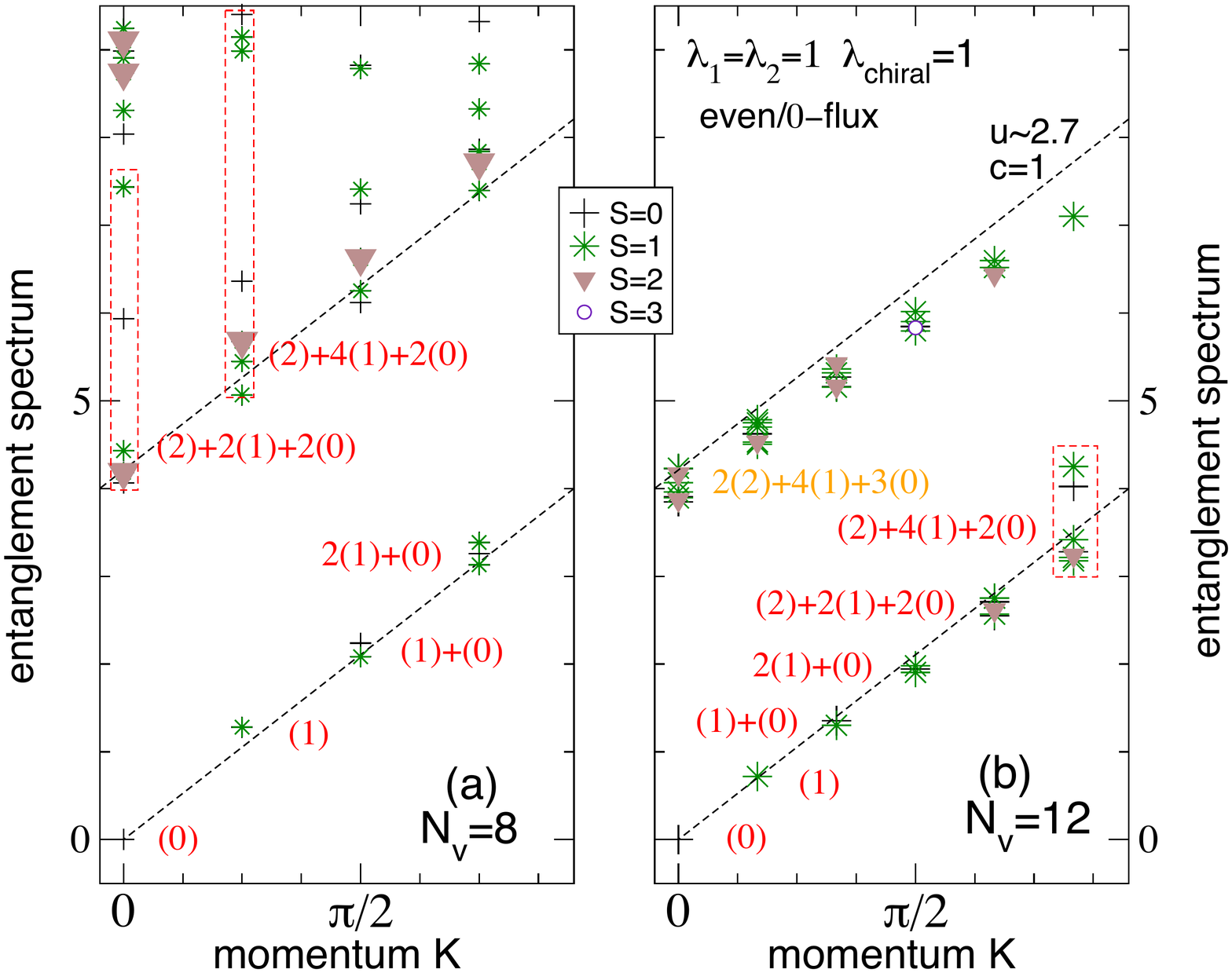}
\includegraphics[width=\columnwidth,angle=0]{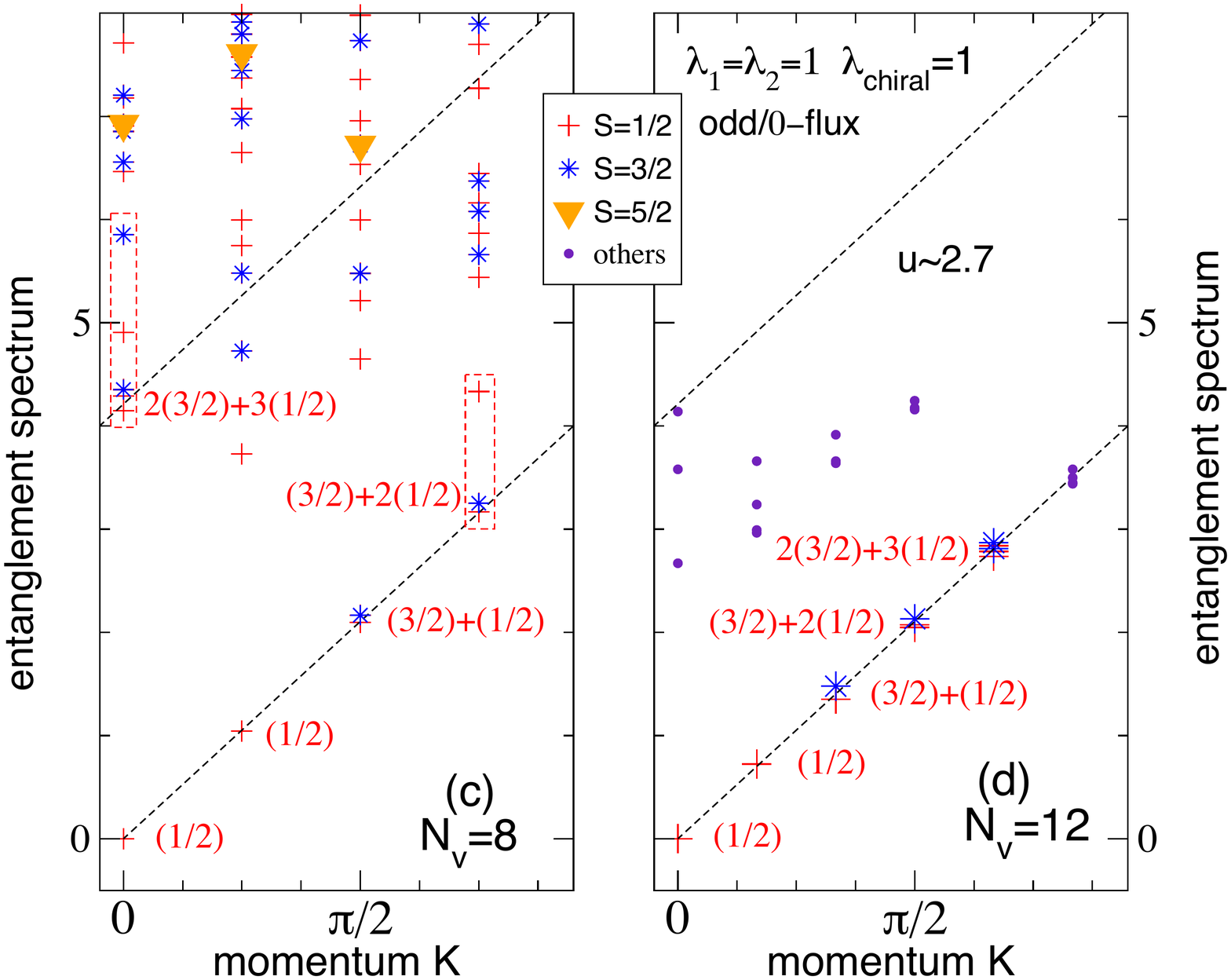}
\caption{Low-energy Entanglement Spectrum obtained on an infinite cylinder of perimeter $N_v$ 
vs momentum along the edge (modulus $\pi$) in even (a,b) and odd (c,d) sectors with 0-flux. The 
GS energy has been subtracted out (to ease the comparison between different sizes). The symbol 
$(S)$ refers to a spin-S multiplet and the correct (slightly incomplete) SU(2)$_1$ counting obtained for each 
quasi-degenerate group of levels -- outlined by boxes when necessary -- is indicated in red (orange). 
(a,c) $N_v=8$ and exact contractions; 
(b,d) $N_v=12$ and approximate contractions done using a $D_c=36$ iMPS.  For each momentum, the lowest 24 quasi-energies are shown. Fits of the linear dispersion in (b) and (d) give the mode velocity $u\sim 2.7$, also consistent 
with the other data in (a) and (c). 
  }
\label{Fig:ES}
\end{center}
\vskip -0.5cm
\end{figure}

{\it Entanglement Spectrum and CFT's} -- The Entanglement Spectrum (ES) is defined as the spectrum of $H_E$.  
It inherits from the spin-singlet character of the PEPS 
SU(2) Kramer degeneracies so that $SU(2)_k$ WZW theories
are the most natural candidate CFT's to describe the low-energy modes. 
We first focus on the two topological sectors, even and odd, in the absence of any vison flux ($f=0$)
and we label the two sectors only by their parity $(p)$.
The corresponding ES on infinite cylinders with $N_v=8$ and $12$ 
are displayed in Fig.~\ref{Fig:ES} as a function of the momentum (modulus $\pi$) 
along the edge of the half-cylinder. 
Low (quasi)energy linearly dispersing chiral modes -- with the same velocity $u$ in all sectors -- 
are clearly seen for both values of $N_v$, in perfect agreement with 
the prediction of $SU(2)_1$ CFT (free bosons or ``Luttinger liquid" at the SU(2)-symmetric point). Indeed, the two spectra in the even and odd 
topological sectors correspond exactly to the two $SU(2)_1$ conformal towers 
associated to each WZW primary field (spin-0 and spin-1/2) and 
its descendants (integer and half-integer spins, respectively) with the right expected degeneracy 
(or quasi-degeneracy) of the 
equally-spaced groups of levels. 
We also observe that the assignment becomes better for increasing cylinder perimeter
and that the odd spectrum is exactly {\it doubly degenerate}. 
 
The low-energy ES can then be represented accurately as ~:
\begin{equation}E\equiv N_v e_0 + e_{\rm topo}  + E_{\rm CFT}\, , 
\label{es}\end{equation} 
where $E_{\rm CFT}$  is the conventional $SU(2)_1$ CFT spectrum~:
\begin{eqnarray}
E_{\rm CFT}(S_z,\{m_n\})=\frac{\pi u}{N_v} (-\frac{c}{24}+ S_z^2 + \sum_{n\ge 1}{m_n n})\, , 
\end{eqnarray}
Above, the central charge is $c=1$. Each state is labelled by $S_z$, $m_1, m_2, m_3, \ldots$  where $m_n\ge 0$ is the occupation number of the harmonic oscillator mode with wave vector $\pi n/N_v$ and there is a harmonic oscillator mode for $n=1,2,3,\ldots$.
The even (odd) sector corresponds to all integer (half-integer) $S_z$ quantum numbers.
All these states can be grouped into SU(2) multiplets.

The non-universal constant $e_0$ in (\ref{es}) is in fact completely fixed by the normalization of the RDMs
and depends only on the chiral mode velocity.
To prove this, we conveniently define the ``partition function" in each topological sector separately 
as~:
\begin{eqnarray} Z^{(p)}(\beta)&=&\tilde{\rm Tr}[\exp{(-\beta H_{E}^{(p)})}]
\label{Zbeta}\\
&=&\exp(-\beta N_v e_0 -\beta e_{\rm topo}) Z^{(p)}_{\rm CFT}(\beta)\, , \nonumber
\end{eqnarray} where 
$p$ stands for the parity ``e" or ``o" and $\beta$ is a dimensionless parameter.
Note that we define here $\tilde{\rm Tr}\equiv \frac{1}{d^{(p)}}{\rm Tr}$ 
which takes into account possible {\it exact} degeneracy of the odd (even) chiral mode (we observe
$d^{(o)}=2$ and $d^{(e)}=1$) and 
\begin{eqnarray}
Z^{(e)}_{\rm CFT}(\beta)=\!\!\sum_{S_z\in\mathbb{N}}\sum_{\{m_n\}}\!\!\exp{(-\beta E_{\rm CFT}(S_z,\{m_n\}))}\, ,\\
Z^{(o)}_{\rm CFT}(\beta)=\!\!\sum_{S_z\in\mathbb{N}}\sum_{\{m_n\}}\!\! \exp{(-\beta E_{\rm CFT}(S_z\!+\!\frac{1}{2},\{m_n\}))}\, ,
\end{eqnarray}
where the second sum runs over all occupations of the harmonic oscillator modes i.e. 
$\sum_{\{m_n\}}\equiv\prod_{n\geq 1}\sum_{m_n\ge 0}$.
We impose that $\tilde{\rm Tr}[\rho^{(p)}]=1$ i.e. $Z^{(p)}(1)=1$ for both $p=e$ and $p=o$. When $N_v\gg \beta u$, one can approximate
\begin{equation}
Z^{(p)}_{\rm CFT}(\beta)\simeq g\exp{(\pi N_v/6 u \beta)}\, ,\label{Zcft}
\end{equation}
where $g$ is a ``boundary degeneracy", $g=1/\sqrt{2}$ here, and the right-hand side is $p$-independent. 
Note that higher order corrections to Eq.~(\ref{Zcft}) are exponentially small, down by a factor of 
$\exp{(-\pi N_v/u \beta)}$. The above result follows from a modular transformation expressing the conformal towers $\chi_s (e^{-\pi u\beta /N_v})$ as a linear combination of $\chi_{s'}(e^{-4\pi N_v/u\beta})$, hence relating the behavior at small $u\beta /N_v$ to the behavior at large $u\beta /N_v$, low temperature, which is easily obtained by considering just a few lowest energy states~\cite{IA_review}. 
Hence, the normalization conditions
imply that 
\begin{eqnarray}
e_0&=&\frac{\pi}{6u}\, ,\label{e0}\\
e_{\rm topo}&=&\ln{g}\label{etopo}\\
&=&-\ln{2}/2 \, ,\nonumber
\end{eqnarray}
which are identical in the even and odd sectors.

{\it Entanglement Entropy} -- 
The Renyi Entanglement Entropies (EE) are defined by $S_{q}^{(p)}=\frac{1}{1-q}\ln{[\tilde{\rm Tr} \{(\rho^{(p)})^q\}]}$.
In the limit $q\rightarrow\infty$, the Renyi EE $S_\infty^{(p)}$ reduces to the ground state energy $E_{\rm gs}^{(p)}$
of the ES in each parity sector.
From the previous analysis, one expects then that, asymptotically for large $N_v$,
\begin{eqnarray}
S_\infty^{(p)}\sim e_0 N_v + e_{\rm topo}  + \frac{\pi u}{N_v} \alpha^{(p)}\, ,
\label{renyi}
\end{eqnarray}
where $e_0$ and $e_{\rm topo}$ are given by Eqs. (\ref{e0}),(\ref{etopo}).
Neglecting subleading terms in Eq. (\ref{Zcft}), one gets $\alpha^{(e)}= -\frac{1}{24}$
and $\alpha^{(o)}= -\frac{1}{24}+\frac{1}{4}$, where the extra $\frac{1}{4}$ is the conformal weight of the
odd sector.

\begin{figure}
\begin{center}
\includegraphics[width=\columnwidth,angle=0]{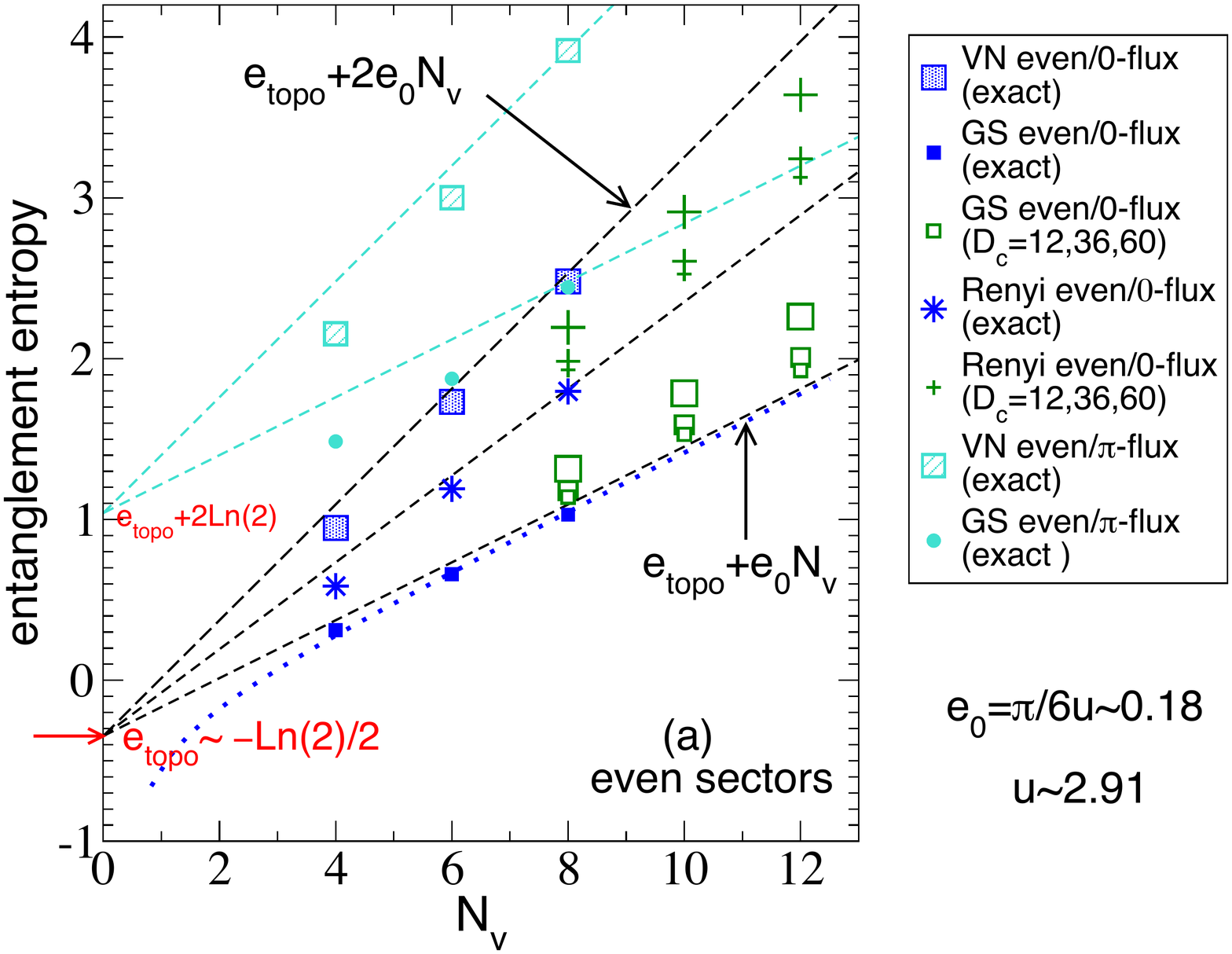}
\includegraphics[width=\columnwidth,angle=0]{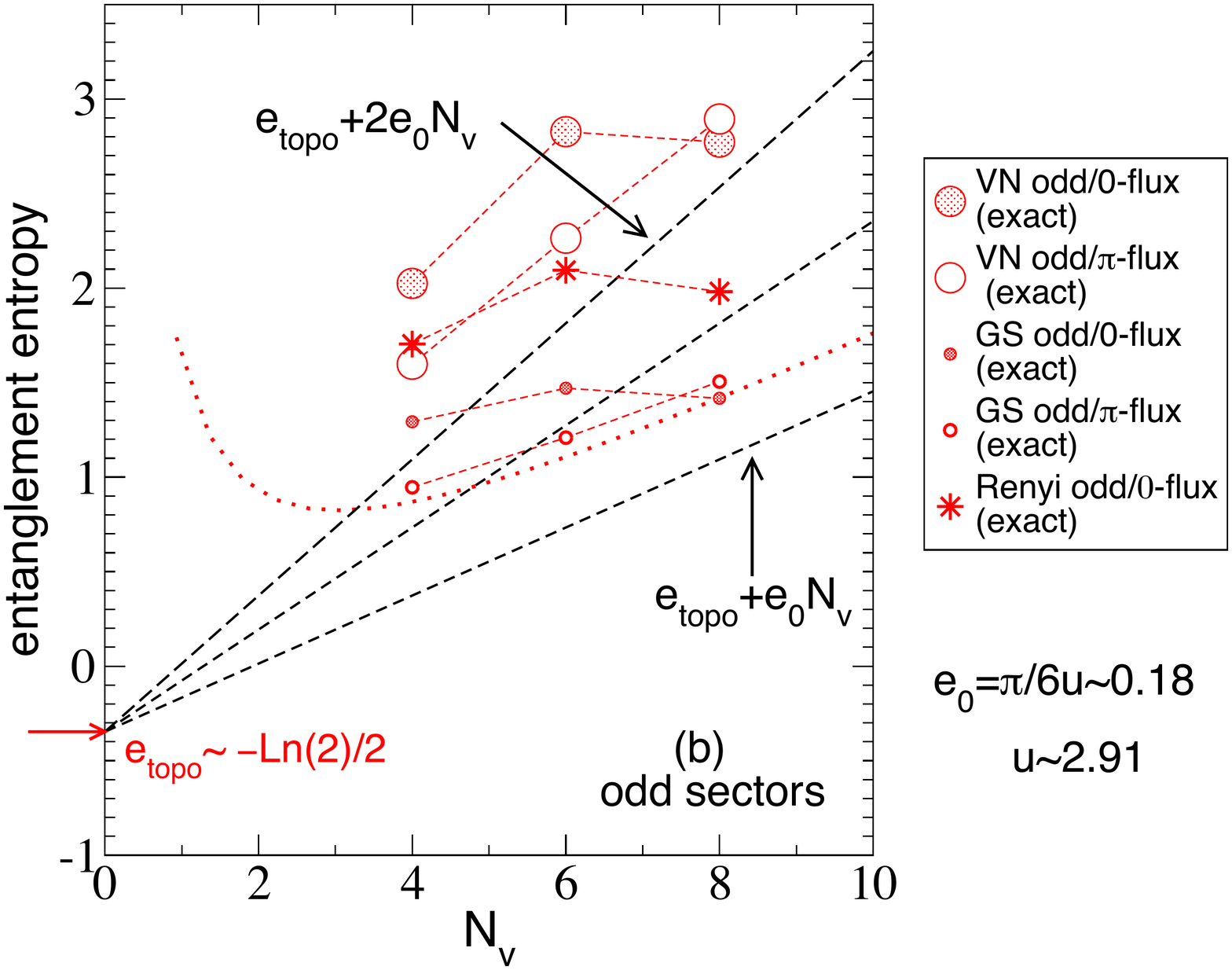}
\caption{VN entanglement entropy, $q=2$ Renyi entropy and ground state of ES (i.e. $q=\infty$ Renyi entropy) vs cylinder perimeter in the topological even sectors (a) and odd sectors (b) obtained from exact contractions. In (a) results from approximate iMPS methods
with $D_c=12, 36, 60$ are also shown (with symbol size decreasing with increasing $D_c$).
The dashed (black) lines are given by fits according to the analytic linear predictions of Eqs.~(\ref{VN}), (\ref{renyi}) and (\ref{renyi_all}) for $q=2$. The extracted constant $e_0$
gives a velocity $u\sim 2.9$ consistent with the mode dispersion in Fig.~\protect\ref{Fig:ES}. The (blue and red) dotted lines are given by Eq.~(\ref{renyi}),
including the  $1/N_v$ corrections which differ in the even and odd sectors. 
In (a) the turquoise dashed lines are the linear behaviors of Eqs.~(\ref{VN}) and (\ref{renyi}) shifted by $2\ln{2}$. 
}
\label{Fig:EE}
\end{center}
\vskip -0.5cm
\end{figure}

In the limit $q\rightarrow 1$, one gets the Von Neumann (VN) EE 
$S_{VN}^{(p)}=-\tilde{\rm Tr} [\rho^{(p)} \ln\rho^{(p)}]$.  Its expected scaling behavior can be obtained from $S_{VN}^{(p)}=S_{\rm thermo}^{(p)}(1)$, where the ``thermodynamic" entropy is defined as $S_{\rm thermo}^{(p)}(\beta)=\beta^2 \frac{\partial F^{(p)}}{\partial \beta}$ with
$F^{(p)}(\beta)=-\frac{1}{\beta} \ln{[Z^{(p)}(\beta)]}$.
It is straightforward to get, when $N_v\gg \beta u$, 
$F^{(p)}(\beta)\simeq(1-\frac{1}{\beta})e_{\rm topo}+(1-\frac{1}{\beta^2})N_ve_0$
and, eventually, 
\begin{eqnarray}
S^{(p)}_{VN}\sim (2 e_0) N_v+ e_{\rm topo} \, .
\label{Eq:VN}
\label{VN}
\end{eqnarray}
where the asymptotic result does not depend on the topological sector. 
It is interesting to see that the coefficient of the ``area law" is twice the one of the $q=\infty$ Renyi EE
(\ref{renyi}) while the topological entropy (i.e. the constant term) is the same.
Note that, strictly speaking, SU(2)$_1$ CFT only gives exponentially small higher order corrections to Eq.~(\ref{Eq:VN}).
However, in application of CFT to the true edge physics, other corrections could arise from irrelevant operators which could be down by only powers of $u /N_v$, rather than being exponentially small.

We now analytically compute the Renyi entropy for general $q$. From its definition and from (\ref{Zbeta}) one gets,
\begin{eqnarray} S_q^{(p)}&=&\frac{1}{1-q}\ln Z^{(p)}(q)\\
&=&\frac{1}{1-q}(-q N_v e_0 -q e_{\rm topo} +\ln Z^{(p)}_{\rm CFT}(q))\, . \nonumber
\end{eqnarray} 
Using (\ref{Zcft}) for $N_v\gg q u$, one gets
\begin{eqnarray}
S_q^{(p)}\sim \frac{q+1}{q}e_0 N_v + e_{\rm topo} \, ,
\label{renyi_all}
\end{eqnarray}
which does not depend on the topological sector and also applies to the previous cases $q=1$ and $q\rightarrow\infty$
(although $1/N_v$ corrections appear in the last case, see (\ref{renyi})).

To compare with the above CFT predictions, we have numerically computed the EEs i) exactly on infinite cylinders for $N_v=4, 6, 8$
and ii) using an approximate iMPS method for $N_v=8, 10, 12$ (only for
$S_{\infty}$ and $S_2$ in the even sector for which we get reliable
results; see Appendix C for details). As shown in Fig.~\ref{Fig:EE}(a),
the numerical results for the even sector
agree very well with the analytic predictions which depend on a single parameter, the mode velocity $u$, estimated independently
from the slope of the chiral mode in Fig.~\ref{Fig:ES}. We note however, that the iMPS EE data depend strongly 
on the MPS bond dimension $D_c$, but in a way which is consistent with the
CFT predictions. Indeed, the $D_c\rightarrow\infty$ limits
of $S_\infty$ and $S_2$, for every circumference $N_v$, seem to coincide with the values given by Eq.~(\ref{renyi_all}) 
(see Appendix C). 
In addition, we see that the  $D_c\rightarrow\infty$ extrapolations of the fitting parameters $e_0(D_c)$ and $e_{\rm topo}(D_c)$ obtained from 
linear fits of $S_\infty(N_v)$  and $S_2(N_v)$
for $8\le N_v\le 12$, independently for each value of $D_c=12,36$ and $60$, are consistent with $e_0\sim 0.18$ and $e_{\rm topo}=-\ln{2}/2$
obtained from the analysis of the exact data for $4\le N_v\le 8$. 
In the odd sector, as seen in Fig.~\ref{Fig:EE}(b), large deviations from the analytic predictions are found for small perimeters 
like $N_v=4,6$. However, the (exact) results for $N_v=8$ become quite close to what is expected from the CFT analysis. 

\begin{figure}
\begin{center}
\includegraphics[width=\columnwidth,angle=0]{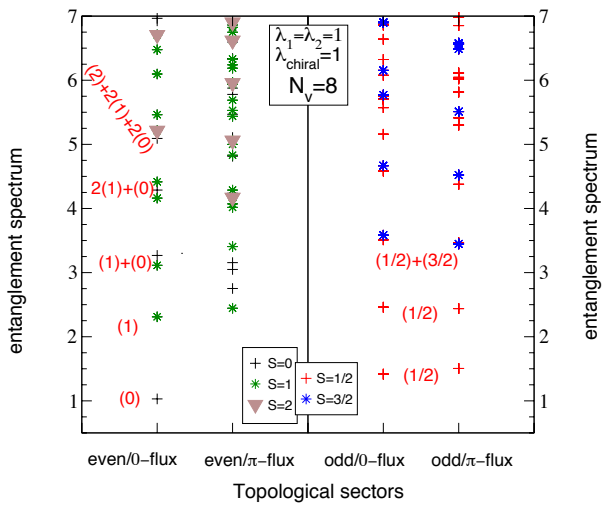}
\caption{Comparison between the 4 low-energy ES in the even (left) and odd (right) sectors with $0$ or $\pi$ vison 
fluxes and for $N_v=8$. The different symbols 
correspond to different spin multiplets and the correct SU(2)$_1$ counting obtained for the low-energy group of levels is 
indicated in red. 
All multiplets of the odd spectra are exactly doubly degenerate. No simple CFT spectrum could be assigned to
the $(e,\pi)$ sector (see text).
  }
\label{Fig:ALL}
\end{center}
\vskip -0.5cm
\end{figure}

{\it $\mathbb Z_2$ Vison flux} -- We now investigate the effect on the ES and the EE of inserting a vison flux ($f=\pi$) inside the cylinder.
As shown in Fig.~\ref{Fig:ALL} for $N_v=8$, the two spectra in the odd sector with and without flux are very similar 
and agree with the half-integer spin sector of SU(2)$_1$.
As a result, the corresponding VN and $q=\infty$ Renyi EE for $f=\pi$ are indeed close to the ones for $f=0$, and not too far from the CFT prediction, as seen in Fig.~\ref{Fig:EE}(b). 
In contrast, we failed to identify 
a simple CFT characterizing the even sector in the presence of a $\mathbb{Z}_2$-flux. By analyzing the VN and $q=\infty$ Renyi EE shown in Fig.~\ref{Fig:EE}(a), we found some similarity between the behaviors for $f=0$ and $f=\pi$ 
up to a global shift. This suggests that the $f=\pi$ spectrum may be described by the product of either i) the CFT odd sector 
by a single free spin (${\rm CFT}_{\rm odd}\otimes\frac{1}{2}$) or ii) the CFT even sector
by two free spins (${\rm CFT}_{\rm even}\otimes\frac{1}{2}\otimes\frac{1}{2}$) so that the boundary degeneracy
i) $g\rightarrow 2g$ or ii) $g\rightarrow 4g$ causing a shift in the EE of i) $\ln{2}$ or ii) $2\ln{2}$. In such scenarii, one expects
either i) (1)+(0) ground states and (1)+(0) first excited states or ii) (1)+(0) ground states and (2)+2(1)+(0) first excited states.
The even/$\pi$-flux low energy spectrum in Fig.~\ref{Fig:ALL} would be more consistent with scenario i) (although there is one 
extra singlet excitation). However, the shift observed in the EE in Fig.~\ref{Fig:EE}(a) is consistent with $2\ln{2}$ which would 
rather point towards ii). We believe however that the system sizes at hand are still too small to enable to reach a definite conclusion. 

{\it Emergent $U(1)$ symmetry and criticality} --
The existence of the 4 topological sectors is directly linked to the $\mathbb{Z}_2$ gauge symmetry~\cite{ran} of the tensor (\ref{Eq:tensor}).
However, a careful (finite size scaling) analysis of the TM's spectrum
(see Appendix B) suggests the latter is gapless, implying long range
(dimer) correlations, typical of the nearest-neighbor RVB state which bears an enlarged $U(1)$ gauge symmetry~\cite{albuquerque,tang}. 
Such features could nevertheless be reconciled by invoking in our case an "emergent" $U(1)$ symmetry. In other words, due to the ``teleportation'' of the
$R_2$ and $I_c$ tensors, the 
chiral PEPS can be viewed as a RVB state with a distribution of singlet bonds extending to 
all distances (always connecting two different sublattices). However, as for the
non-chiral RVB state of Ref.~\onlinecite{wang}, this distribution falls off exponentially fast with the bond length. 
Therefore, for very wide cylinders, the probability to find a singlet bond winding around the cylinder is exponentially suppressed and
the (chiral or non-chiral) RVB states behave effectively as ``short-range" RVB states possessing enlarged $U(1)$ gauge symmetry. 

{\it Outlook} -- 
In this paper, we investigated in details the nature of the ES (reflecting the edge physics) of a simple chiral critical spin liquid
given by a $D=3$ PEPS. We numerically observed that the low-energy ES follows the 
exact counting of a SU(2)$_1$ Wess-Zumino-Witten CFT. 
We derived analytically the expected behavior of the Renyi and Von Neumann Entanglement Entropies assuming a {\it perfect} 
CFT spectrum. A comparison to numerical results confirms the expected topological term of $-\ln{2}/2$, as in the
half-filled (bosonic) Laughlin state. We also provide some evidence that the bulk is critical. Therefore, we argue that 
our family of PEPS physically describes a boundary between a trivial phase and a (gapped) chiral topological phase from
which it inherits the edge states.

{\bf Acknowledgment} --  This project is supported by the NQPTP
ANR-0406-01 grant (French Research Council), the ERC Starting Grant
No.~636201 WASCOSYS and CIFAR (IA).  The numerical computations have been achieved at the
CALMIP EOS Supercomputer (Toulouse) and the JARA-HPC cluster
(Aachen/Juelich) through grant jara0092 and jara0111.  I.A. would like to thank Laboratoire de Physique Th\'eorique, Universit\'e de Toulouse 
for hospitality during the time some of this research was performed. 
We acknowledge
useful conversations with Ignacio Cirac, J\'er\^ome Dubail, Beno{\^\i}t Estienne, Pierre
Pujol, Ying Ran, Nicolas Regnault and Guifre Vidal.

\begin{appendices}

\section{Appendix A~: $D=3$ PEPS ansatz}

\begin{figure}
\begin{center}
\includegraphics[width=8cm,angle=0]{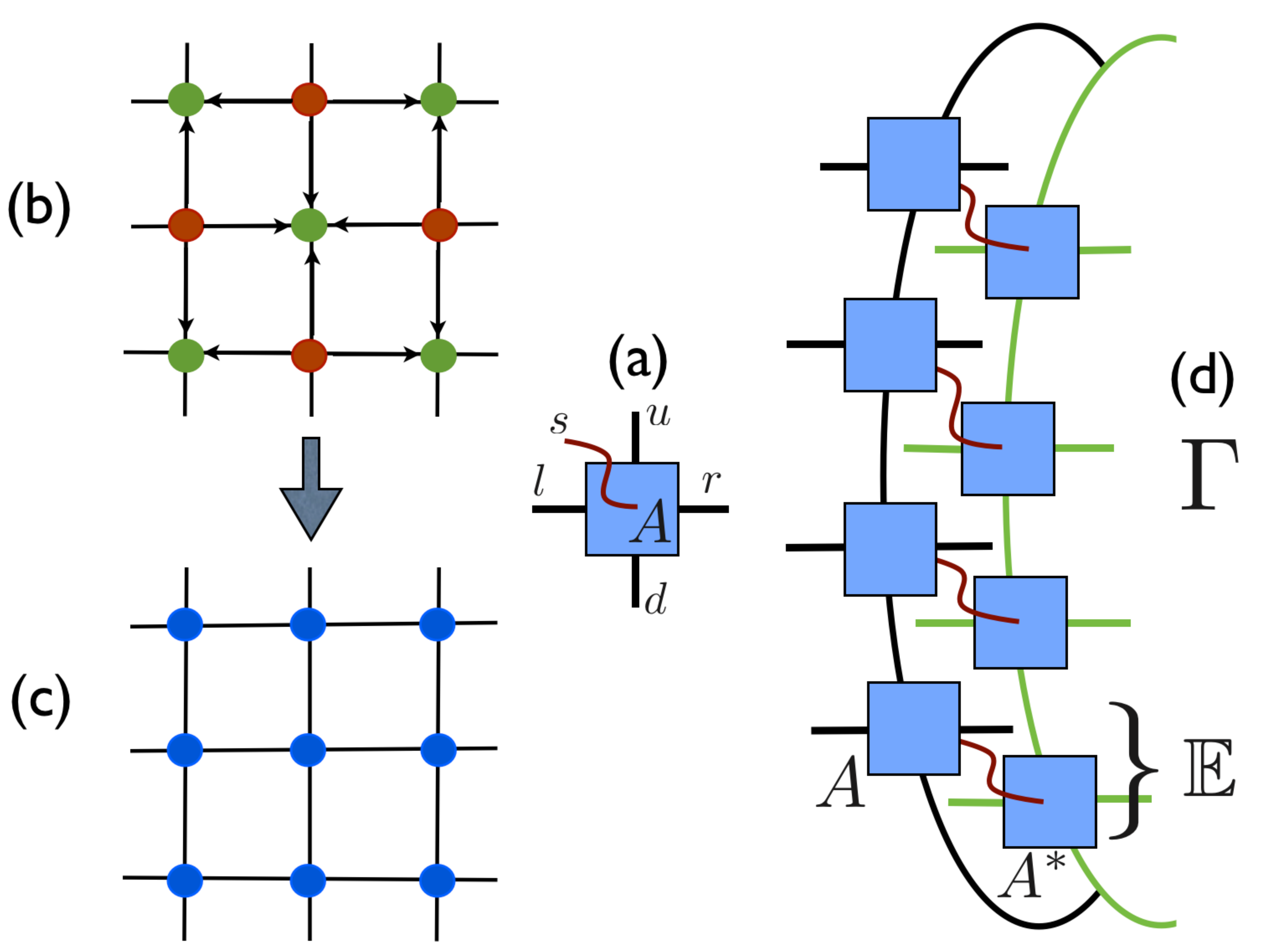}
\caption{(a) Local tensor $A$.
(b) Square lattice with singlet bonds oriented from sublattice A to sublattice B. (c) Under a 180-degrees spin rotation on 
e.g. the B sublattice, oriented singlets are transformed into symmetric 
$|\uparrow\uparrow\rangle+|\downarrow\downarrow\rangle$ maximally entangled pair states. (d) Transfer matrix on a periodic ring. 
  }
\label{Fig:lattice}
\end{center}
\vskip -0.5cm
\end{figure}

The on-site tensor ansatz is 
a rank-5 tensor $A^s_{lurd}$ assigned to each site of a square lattice. Here, the index $s=0,1$ stands for the local physical spin-1/2 degrees of freedom (of dimension $d=2$) and the subscript indices $l,u,r,d$ label the virtual states on the bonds as shown in Fig.~\ref{Fig:lattice}(a).
In order to construct a generalized spin-singlet RVB wavefunction, we assume that the virtual states belong to the $1/2\oplus 0$ spin representation of dimension $D=3$ (the virtual states are labelled by 0, 1 and 2 and carry $
S_z=1/2,-1/2$ and $0$, respectively)~\cite{norbert,didier}. Basically, we can write a general spin-1/2 ansatz wave function on a periodic manifold of $N$ sites as $|\Psi\rangle = \sum_{s_1,s_2\dots i_N}c_{s_1,s_2\dots s_N}|s_1,s_2\dots s_N\rangle$, and $c_{s_1,s_2\dots s_N}={\rm Contract}[A^{s_1}A^{s_2}\dots A^{s_N}]$ where all tensors 
share the same bond variables with their 4 neighbors and ``Contract'' means that one sums up over all bond variables of the 2d tensor 
network. 
Whenever open boundaries are present, one needs to assign the values of the non-contracted virtual indices at the boundaries to fully characterize 
the PEPS wavefunction. Note that the precise sign structure of the RVB state depends on the convention 
to orient singlets: hereafter singlets are oriented from one sublattice (A) to the other sublattice (B) as shown in Fig.~\ref{Fig:lattice}(b).  After  a 180-degrees spin-rotation on the B sites
(see Fig.~\ref{Fig:lattice}(c)) the RVB state takes the form of a translationally invariant PEPS with {\it the same} tensor on both sublattices.

Let us write $A = R+ i I$, where $R$ and $I$ are two real tensors.  We impose that  $R$ and $I$
are, under point group operations, of $B_1$ ($d_{x^2-y^2}$) and $B_2$ ($d_{xy}$) symmetries, respectively.
These constraints together with the additional ones imposed by the singlet character of the PEPS fully determine the
tensor A~\cite{chiral_SL}.
We recall here the form of the $R$ and $I$ (real) components:
\begin{eqnarray}
R^{s'}_{2s22}&=& \lambda_1^{ss'}   ,  \\
R^{s'}_{222s}&=& \lambda_1^{ss'}  ,   \\
R^{s'}_{22s2}&=& - \lambda_1^{ss'}  ,  \\
R^{s'}_{s222}&=&- \lambda_1^{ss'}  ,   
\end{eqnarray}
where spin SU(2)-invariance implies that,
\begin{equation}
\lambda_1^{ss'}= \delta_{ss'}\lambda_1 .
\end{equation}
These tensor elements correspond exactly to those of the NN RVB wavefunction.  
The tensor elements of $I$ with the same indices identically vanish by symmetry.
Non-zero tensor elements that correspond to quantum teleportation via diagonal bonds
are given by~:
\begin{eqnarray}
R^{s'}_{s{\bar s}s2}&=& \lambda_0^{ss'}   ,  \\
R^{s'}_{s2s{\bar s}}&=& \lambda_0 ^{ss'}  ,  \\
R^{s'}_{2s{\bar s}s}&=& - \lambda_0^{ss'}   ,  \\
R^{s'}_{{\bar s}s2s}&=& - \lambda_0^{ss'}   ,  
\end{eqnarray}
where $\lambda_0^{ss'}\in {\mathbb R}$. One also gets (grouping the $R$ and $I$ tensors),
\begin{eqnarray}
A_{ss{\bar s}2}^{s'}&=& \lambda_2^{ss'}  + i \lambda_{\rm chiral}^{ss'}  ,  \\
A_{{\bar s}2ss}^{s'}&=& \lambda_2^{ss'}  + i \lambda_{\rm chiral}^{ss'}  ,  \\
A_{s2{\bar s}s}^{s'}&=& \lambda_2^{ss'}  - i \lambda_{\rm chiral}^{ss'}  ,  \\
A_{{\bar s}ss2}^{s'}&=& \lambda_2^{ss'}  - i \lambda_{\rm chiral}^{ss'}  ,  \\
A_{ss2{\bar s}}^{s'}&=& -\lambda_2^{ss'}  + i \lambda_{\rm chiral}^{ss'}  ,  \\
A_{2{\bar s}ss}^{s'}&=& -\lambda_2^{ss'}  + i \lambda_{\rm chiral}^{ss'}  ,  \\
A_{s{\bar s}2s}^{s'}&=& -\lambda_2^{ss'}  - i \lambda_{\rm chiral}^{ss'}  ,  \\
A_{2ss{\bar s}}^{s'}&=& -\lambda_2^{ss'}  - i \lambda_{\rm chiral}^{ss'}  ,  
\end{eqnarray}
where  $\lambda_2^{ss'}, \lambda_{\rm chiral}^{ss'}\in {\mathbb R}$ are independent constants
and $\bar s$ is the time-reversed of the spin $s$.

Finally, enforcing  the invariance under SU(2) spin-rotations (singlet character of the PEPS) one finds that
$\lambda_2^{ss'}$, $\lambda_{\rm chiral}^{ss'}$ and $\lambda_0^{ss'}$ are diagonal in the spin indices and
odd under spin inversion (we thank Dr. Shuo Yang for pointing out a misprint 
in Eq.~(A20) of Ref.~\onlinecite{chiral_SL}),
 \begin{eqnarray}
\lambda_2^{ss'}&=&\delta_{ss'} (-1)^s \lambda_2, \label{misprint} \\
\lambda_{\rm chiral}^{ss'}&=&\delta_{ss'} (-1)^s \lambda_{\rm chiral}, \\
\lambda_0^{ss'}&=& \delta_{ss'} (-1)^s\lambda_0.
\end{eqnarray}
Note that spin rotation invariance leads also to the relation,
\begin{equation}
\lambda_0=2\lambda_2.
\end{equation}

\section{Appendix B~: Transfer matrix and topological sectors} 
We have put the PEPS wavefunction on a (horizontal) cylinder of circumference 
$N_v$ ($=4, 6, 8$) and length $N_h\rightarrow\infty$. 
The normalization  takes the form
$\langle\Psi|\Psi\rangle = V_{\rm left} \Gamma^{N_h} V_{\rm right}$  where $\Gamma$ is the $D^{2N_v}\times D^{2N_v}$ transfer matrices (TM) and $V_{\rm left}$ and $V_{\rm right}$ stand 
for left and right vectors (of dimension $D^{2N_v}$, $D=3$) defining the boundary conditions. 
The block structure of the transfer matrix and the leading eigenvalues provide key informations on the nature of the state~\cite{norbert2}
such as its topological properties, its correlation lengths, etc... 
To construct $\Gamma$ one first defines a local rank-4 ${\mathbb E}$ tensor by contracting the physical index of two superposed
$A$ tensors, namely 
\begin{equation}
\label{eq:E_def}
{\mathbb E}_{LURD}= \sum_sA^s_{lurd} (A^s_{l' u' r' d'})^*
\end{equation}
where 
the indices of $\mathbb E$ combine the indices of the superposed bonds of the top (bra) and bottom (ket) $A$ tensors.
Then, one builds a (vertical) 
periodic array of $N_v$ such $\mathbb E$ tensors, contracting over the (vertical) bond indices as shown in Fig.~\ref{Fig:lattice}(d). The result can indeed be viewed 
as a $D^{2N_v}\times D^{2N_v}$ matrix 
connecting the (non-contracted) bra and ket virtual indices on the left of the ring to the ones on its right. Due to
$\mathbb{Z}_2$ symmetry the transfer matrix contains four disconnected blocks associated to the topological sectors.

\begin{figure}
\begin{center}
\includegraphics[width=8cm,angle=0]{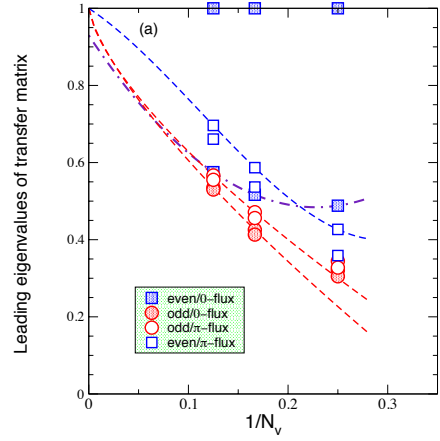}
\caption{ Finite size scaling of the leading eigenvalues of the transfer matrix vs the inverse of the cylinder perimeter.
For each sector (corresponding to a given symbol) the two largest eigenvalues are shown. Dashed lines are guide to the eyes.}
\label{Fig:weights}
\end{center}
\vskip -0.5cm
\end{figure}

We have plotted the first two leading eigenvalues (normalized to the absolute largest) of the transfer matrix in each block as a function of the inverse circumference in Fig.~\ref{Fig:weights}. As expected the leading eigenvalues in all four sectors
seem to approach the same limit ($=1$) for infinite circumference. The gaps defined by the difference between the two leading
eigenvalues of each sector seem also to vanish. However, we note that the finite size gap in the even/$0$-flux sector is much larger than in the other sectors.

\section{Appendix C~: iMPS method}

In this appendix, we give the details of the iMPS method
used to determine the entanglement properties of the chiral PEPS. We will
discuss the general case of a non-hermitian transfer operator (i.e., with
different left and right fixed points), even though the examples
considered have a Hermitian transfer operator by virtue of their
underlying $d$-wave symmetry.

First, we use infinite tranlational invariant MPS (iMPS) to approximate
the (left and right) fixed point of the transfer operator with an infinite
translational invariant MPS, i.e., an ansatz of the form (w.l.o.g.\ we
consider the right fixed point)
\[
\sum_{\{L_k\}}
    (\cdots M^{L_{-1}} M^{L_0} M^{L_1}\cdots)\,
    \vert\dots,L_{-1},L_0,L_1,\dots\rangle\ ,
\]
where $L_{k}$, $k\in\mathbb Z$, label the (left-pointing) indices of
(right) fixed point, cf.~Appendix~B, and the $M^{L}$ are
$D_c\times D_c$ matrices.  To this end, we start with some initial tensor
$M^L$. Applying the transfer operator [cf.~Eq.~(\ref{eq:E_def})] yields a
new iMPS with matrices
\[
\tilde{M}^L = \sum_{L',R} \mathbb E_{LURD}\,M^{L'}\delta_{L'R}
\]
with bond dimension $D^2D_c=9D_c$. We now truncate the bond dimension by bringing
the iMPS into the canonical form of Ref.~\cite{vidal:iTEBD} (i.e., with weights
$\cdots\Lambda \Gamma^{L_i}\Lambda\Gamma^{L_{i+1}}\cdots$, with $\Lambda$
a diagonal matrix holding the Schmidt coefficients of the bipartition),
and discarding all but the $D_c$ largest Schmidt coefficients.

\begin{figure}[t]
\begin{center}
\includegraphics[width=\columnwidth,angle=0]{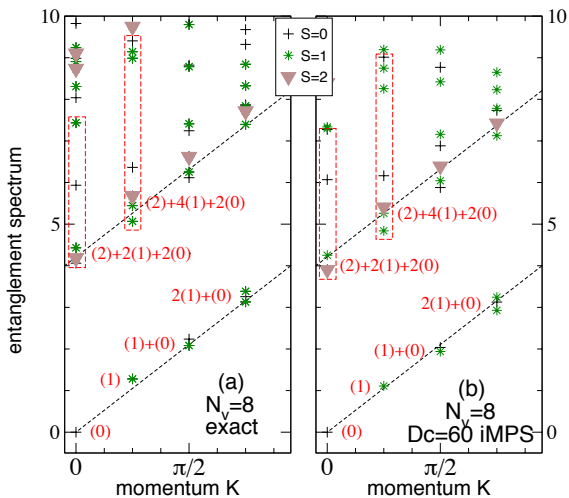}
\caption{Comparison of the low-energy Entanglement Spectrum on an infinite cylinder of perimeter $N_v=8$ 
 in the even/0-flux sector computed by (a) exact contractions and (b) $D_c=60$ iMPS method. 
As in Fig.~\protect\ref{Fig:ES} the GS energy has been subtracted and same symbols are used. 
In (b),  the lowest 24 quasi-energies are shown for each momentum.
 The linear dispersion (dashed lines) correspond to ta mode velocity $u\sim 2.7$.  }
\label{Fig:ES_comparison}
\end{center}
\vskip -0.5cm
\end{figure}

In order to extract the entanglement spectrum from the right and left
fixed point iMPS $M^L$ and $N^R$, note first that the spectrum of
$\rho=\sqrt{(\sigma_R)^\top}\sigma_L\sqrt{(\sigma_R)^\top}$ equals the
spectrum  of $\tau=\sigma_R^\top\sigma_L$, and their eigenvectors share
the same quantum numbers; we use the latter form as it does not
require taking square roots (which cannot be done easily for Matrix
Product Operators~\cite{delascuevas:purifications}). We build $\tau$ on a
cylinder of circumference $N_v$ by putting the fixed point iMPS on a ring
of $N_v$
sites and appropriately contracting indices, so that $\tau$ is expressed
as a Matrix Product Operator (MPO) with tensors $\sum_{s} M^{ls}N^{sr'}$.
We can now use a Krylov method, where the MPO is applied to a vector by
sequential contraction (cf.~Appendix B) to determine the spectrum of
$\tau$, resolved by symmetry sectors and momentum.  As the largest object
which needs to be stored is of the size of the eigenvectors of $\rho$
(rather than of $\rho$ itself), this allows us to go to significantly
larger system sizes.

\begin{figure}[t]
\begin{center}
\includegraphics[width=\columnwidth,angle=0]{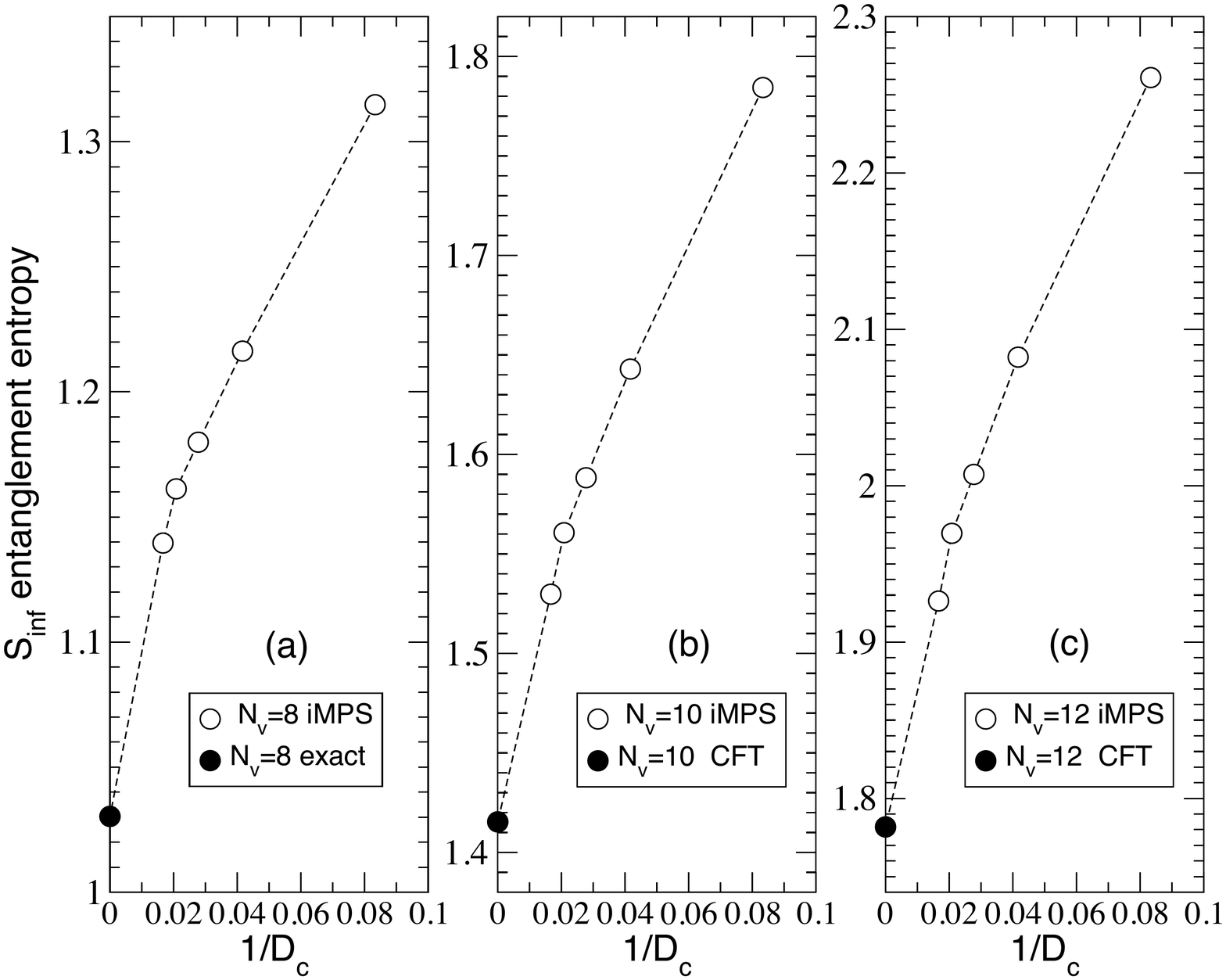}
\includegraphics[width=\columnwidth,angle=0]{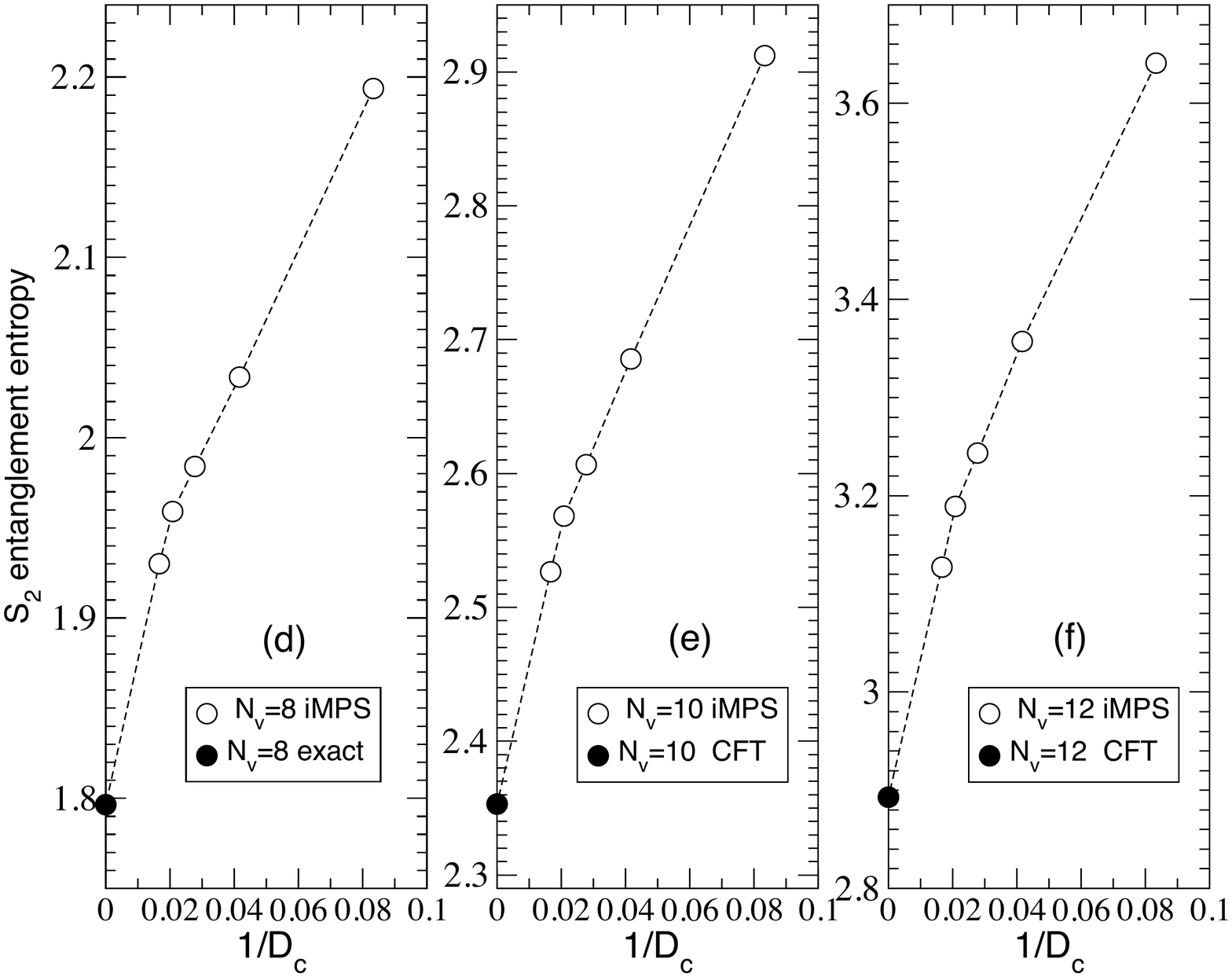}
\caption{Scaling of the  $S_{\infty}$ (GS energy of the ES) and $S_2$ Renyi EE computed with iPEPS vs $1/D_c$ for $N_v=8$ (a,d), $N_v=10$ (b,e) and 
$N_v=12$ (c,f). The exact value (a,d) and the predictions from CFT (b,c,e,f) are also shown for comparison at $1/D_c=0$.}
\label{Fig:EE_vsD}
\end{center}
\vskip -0.5cm
\end{figure}

To determine the entanglement entropies in Fig.~\ref{Fig:EE} from the iMPS
simulations, we have used the leading terms in the entanglement spectrum
as determined by the Krylov method (here, $24$ eigenvalues per momentum):
$S_\infty$ is determined by the leading eigenvalue alone, while
$S_2$ can be approximated using the leading eigenvalues.  We find that
for $S_2$, the data is fully converged in the number of eigenvalues and
only needs to be extrapolated in $D_c$. On the other hand, we have
also observed that e.g. for $S_{VN}$, the convergence in the number of
eigenvalues is significantly slower, rendering a reliable extrapolation
very difficult.

\begin{figure}[t]
\begin{center}
\includegraphics[width=\columnwidth,angle=0]{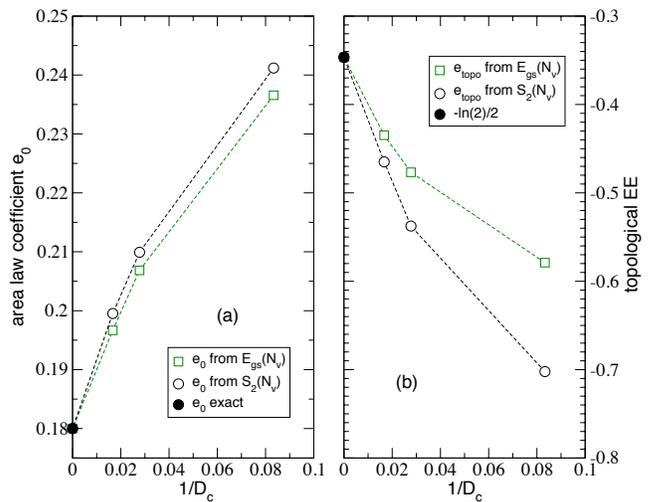}
\caption{Parameters $e_0$ (a) and $e_{\rm topo}$ (b)  vs $1/D_c$ obtained by linear fits of $S_\infty(N_v)$  and $S_2(N_v)$
($8\le N_v\le 12$) independently for each value of $D_c=12,36$ and $60$. At $1/D_c=0$ we report the corresponding values obtained from the analysis of the
exact EEs on infinite cylinders with $N_v=4,6,8$.  
}
\label{Fig:fits_vsD}
\end{center}
\vskip -0.5cm
\end{figure}

It should be noted that since the construction involves taking a solution
obtained for an \emph{infinite} system as an ansatz for a \emph{finite}
system without further optimization, we cannot guarantee the ansatz to
exactly reproduce the exact data for finite $N_v$ even as
$D_\mathrm{c}\rightarrow\infty$.  On the other hand, we observe that the
entanglement spectrum for $N_v=8$ and $D_c=60$ is very close to the exact data 
as seen in Fig.~\ref{Fig:ES_comparison}.  In Fig.~\ref{Fig:EE_vsD} we plot $S_\infty$
and $S_2$ vs $1/D_c$, for each circumference $N_v=8$, $10$ and $12$. We
observe that, although the EE strongly depend on $D_c$,  the
$D_c\rightarrow$ limits of $S_\infty$ and $S_2$ are consistent with the
exact value for $N_v=8$ or the values given by Eq.~(\ref{renyi_all}) for
$N_v=10$ and $12$.  In Fig.~\ref{Fig:fits_vsD} we plot the fitting
parameters $e_0(D_c)$ and $e_{\rm topo}(D_c)$ obtained from linear fits of
$S_\infty(N_v)$  and $S_2(N_v)$ for $8\le N_v\le 12$, independently for
each value of $D_c=12$, $36$ and $60$.  Again, the $D_c\rightarrow$ limits of
$e_0(D_c)$ and $e_{\rm topo}(D_c)$ are consistent with $e_0\sim 0.18$ and
$e_{\rm topo}=-\ln{2}/2$ obtained from the analysis of the exact data for
$4\le N_v\le 8$.  Incidently, since $e_0$ is directly linked to the mode
velocity $u\propto 1/e_0$, we then expect that iMPS gives a 
{\it smaller} velocity $u$ of the chiral mode, which we estimate to deviate when $D_c=60$ by less than 10$\%$ from the
exact calculation, as shown in Fig.~\ref{Fig:ES_comparison} for $N_v=8$.

\begin{figure}[t]
\includegraphics[width=0.95\columnwidth]{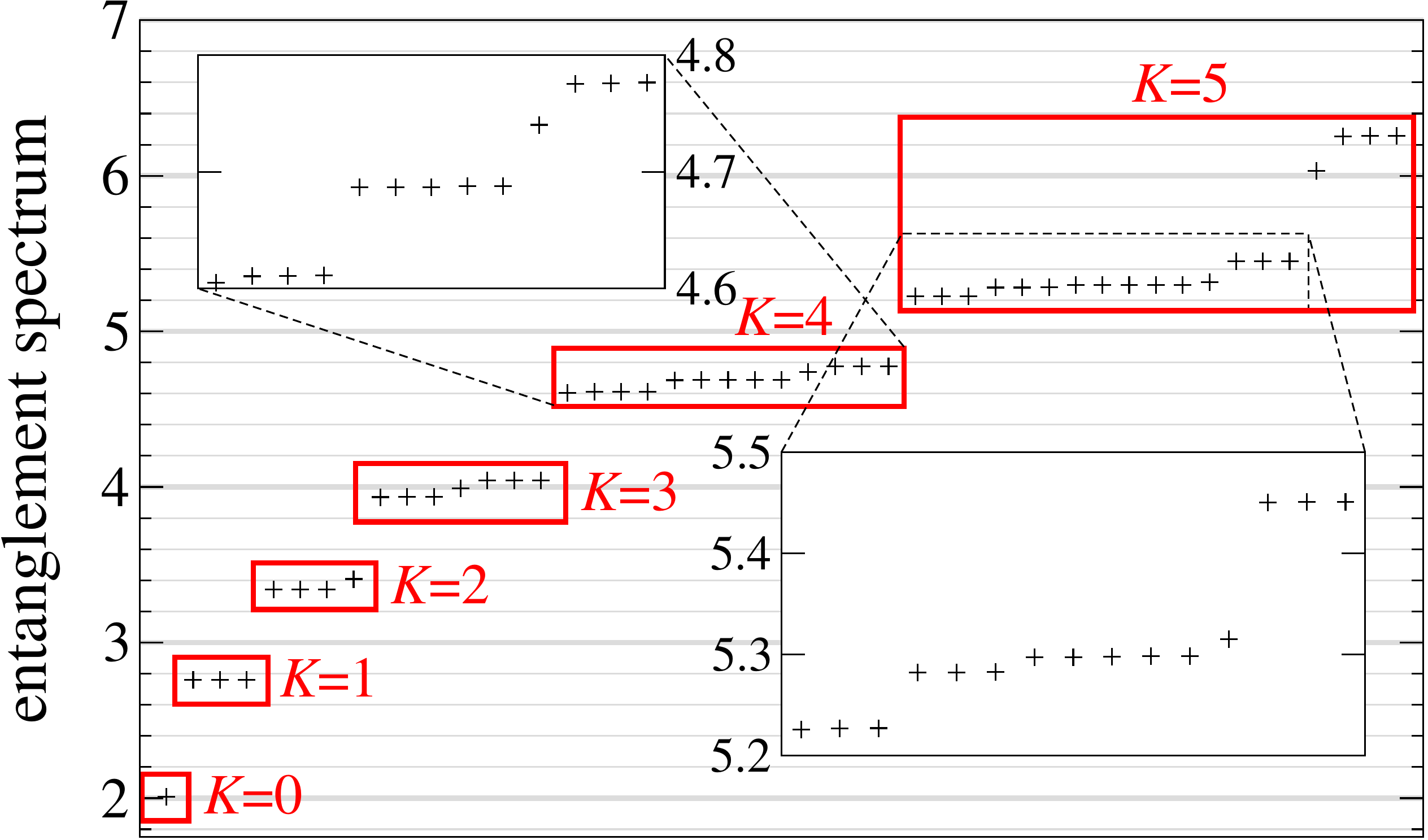}
\caption{
\label{fig:imps_multiplets}
Entanglement spectrum obtained by $D_c=36$ iMPS for $N_v=12$, sorted by
momentum $K$ (in units of $2\pi/12$, and $\mathrm{mod}\,\pi$); the insets
show zooms into the areas marked with dashed boxes.  One can clearly
recognize the degeneracies corresponding to the spin multiplet structure
$(0)$, $(1)$, $(1)+(0)$, $2(1)+(0)$, $(2)+2(1)+2(0)$, $(2)+4(1)+2(0)$.
}
\end{figure}

While we have not enforced $SU(2)$ symmetry when approximating the fixed
point by an iMPS (the only symmetry enforced by the ansatz is
translation), the spin multiplets are nevertheless very well resolved in
the degeneracies of the entanglement spectrum (at least in the even sector) and can be used to infer the
multiplet structure without having to measure the spin, as shown in
Fig.~\ref{fig:imps_multiplets}.

Let us note that an alternative way to approximate the $S_2$ Renyi
entropy, as well as other quantities which can be expressed as polynomials
in the left and right boundaries $\sigma^L$ and $\sigma^R$ (in particular
$q$-Renyi entropies for integer $q$ and the momentum
polarization, including combinations thereof with the even parity
projector, but also the normalization, or half-integer Renyi entropies if
$\sigma^L=\bar\sigma^R$) is to write down the corresponding tensor network for
a system on a cylinder of circumference $N_v$ and to subsequently contract
it in the ``transversal'' direction, i.e., around the cylinder, in which
it is translationally invariant.  This leads to an expression of the form
$\mathrm{tr}(M^{N_v})$ which can be computed for arbitrary $N_v$, either
exactly (by taking matrix powers or diagonalizing $M$), or approximately
by again using a Krylov method to determine the leading eigenvalues of
$M$, which can then be used to approximate $\mathrm{tr}(M^{N_v})$; in
particular, the leading behavior is determined by the largest eigenvalue.
Let us note, however, that this method yields an approximation of the
actual behavior by a sum of exponentials, and thus in order to extract the
true scaling, a careful extrapolation in the relevant truncation
parameters is required.

\end{appendices}

\end{document}